\documentstyle[aps,prd,epsf]{revtex}                                  

\begin{document}

\title{The Preheating - Gravitational Wave Correspondence:~I}

\author{Bruce Bassett\thanks{email: {\sc bruce@stardust.sissa.it}}}

\address{International School for Advanced Studies, SISSA/ISAS, Via 
Beirut 2-4, Trieste 34014, Italy \\
Department of Mathematics and Applied Mathematics, University of 
Cape Town, Rondebosch, 7700, South Africa} 
\date{\today}
\maketitle
\begin{abstract}
The gravitational wave equations form a parametric 
resonance system during oscillatory reheating after inflation, when cast 
in terms of the electric and magnetic parts of the Weyl tensor. This is in 
direct  analogy with preheating. For chaotic inflation with a quadratic 
potential, this analogy is exact. The resulting amplification of the 
gravitational wave  spectrum begins in the broad resonance regime for 
chaotic inflation but quickly moves towards narrow resonance. This 
increases the amplitude and breaks the scale invariance 
of the tensor spectrum generated during inflation, an effect which may be 
detectable. The parametric amplification is absent, however, in 1st-order 
phase transition models and  ``warm" inflationary models with continuous 
entropy production.
\end{abstract}
\section{Introduction}
Standard inflation ends with the universe in a near-vacuum state  
\& hence reheating the universe successfully  is a 
crucial  requirement of  inflationary scenarios. A full theory of this 
phoenix-like  revival from the remains of the inflaton field remains 
elusive given the  complexities of non-equilibrium field theory on curved 
spacetime,  although the  so-called elementary theory has been known  for 
over a  decade and a  half  \cite{DL,Brand,linde90}. However, whereas 
previously the 
importance of  parametric resonance was recognised \cite{Brand,TYB94}, it 
was not studied in the broad-resonance realm. It  
was essentially the move into this non-perturbative  regime - 
{\em preheating} -  which lead to a radical  paradigm shift 
\cite{KLS94}, demonstrating
that inflation may give rise to nonlinear fluctuations of sufficient 
magnitude to  restore 
symmetries \cite{KLS95} and  offer a way of gracefully exiting 
inflation \cite{devegads}. In addition, it has forced the study of 
backreaction phenomena \cite{KLS94,Boyan} using nonperturbative 
methods, and the effects of scattering during the broad 
resonance regime \cite{PR96,tkachev}.

However, the implications of oscillatory reheating (through a second 
order phase transition) and the resulting large quantum fluctuations for 
local  curvature (metric) fluctuations has been largely unexplored until 
now, and limited to scalar modes  
\cite{kodama1,NT96,ml96}. The evolution of  the 
tensor (gravitational wave) spectrum has essentially been completely 
ignored, apart  from a recent study of gravitational 
bremstrahlung generated  through the interactions of the large quantum 
fluctuations in the inflaton and decay-product fields \cite{tkachevgw}. 
Gravitational wave  production during the 
bubble-wall collisions of a first order phase transition have been 
rather more 
studied \cite{KKT94}. However, the gravitational waves produced  in 
these mechanisms are  backreaction phenomena,
since they are due to the scalar field fluctuations, rather than  
the background zero mode evolution itself. 

In this paper we wish to demonstrate that there exists such an 
amplification of gravitational waves, essentially  due to the oscillation of 
the  zero mode of the inflaton during reheating. This is in addition to any 
gravitational  bremstrahlung that may be produced by the associated 
scalar fluctuations. Further, we show that a direct analogy 
exists in the treatment of preheating and gravitational wave evolution 
at the end of inflation if this occurs via a second order phase 
transition, as we shall assume here. Indeed, both are  
governed by (approximately) Floquet systems \footnote{A 
Floquet system is any set of linear ODE's with periodic coefficients.
The solutions of such systems are characterized by resonance bands of 
exponentially growing modes, indexed by the momentum $k$.}. In the case of 
the quadratic 
potential, $V(\phi) = \frac{1}{2}m_{\phi}^2 \phi^2$, both preheating and 
gravitational wave amplification are initially well approximated by the 
Mathieu  equation, hence the correspondence of the title. This ``duality" is 
exhibited using  the covariant Maxwell-Weyl form of the Einstein field 
equations (see  section \ref{sec:tensor} and e.g. \cite{elli71}), and is 
partially   hidden in the Bardeen formalism \cite{pretend2}.  This 
demonstration of  gravitational wave amplification should be considered 
in the light of the  controversy surrounding tensor amplification during 
instantaneous phase  transitions. In these models,   two 
different cosmologies such as de Sitter  and radiation 
Friedmann-Lema\^{\i}tre-Robertson-Walker ({\sc flrw}) are matched 
across a single 3-space through appropriate junction conditions.
Such discussions  are indeed rather technical and subtle, dealing with 
discontinuities in  the Einstein field equations, and do not model the 
actual physics of reheating. Further, there is as yet no absolute clarity 
as to the possibility of tensor amplification in these models, with 
claims both for \cite{gris} and against  \cite{deru,martin} amplification. 

Now, in the covariant approach to cosmology (see e.g. 
\cite{elli71,roy,DBE96}), a fundamental vector is 
the four velocity of the fluid, $u^a$, which is used to perform a $3+1$ 
splitting of spacetime. The expansion, $\Theta$, is then  defined via: 
\begin{equation}
\Theta = u^a{}_{;a}
\label{eq:expand}  
\end{equation}     
where ; denotes the covariant derivative. In {\sc flrw} universes, $\Theta =
3\dot{S}/S = 3H$, where S is the scale factor and H is the Hubble
constant. In the case of a minimally coupled inflaton field $\phi$, we may 
choose our 
$u^a$ to be proportional to the timelike vector field orthogonal to the 
surfaces of constant $\phi$, i.e. proportional to the covariant derivative 
$\phi_{;a}$ \cite{BED92}, with $h_{ab} = g_{ab} + 
u_a u_b$ the projection tensor into the orthogonal 3-spaces. The 
convective time derivative is  defined as $\dot{Q}^{a..b} \equiv 
Q^{a..b}{}_{;c}u^c$, for arbitrary $Q^{a..b}$, i.e. it is the covariant 
derivative projected along the four velocity. This coincides with the 
usual time derivative in {\sc flrw} spacetimes. 
The evolution equation for the inflaton $\phi$, with effective potential
$V_{\mbox{eff}}(\phi)$ \footnote{From here on we drop the {\em $\mbox{eff}$} 
subscript from $V(\phi)$. It is implicit.}, is given by: \begin{equation}
\ddot{\phi}   +  \Theta \dot{\phi} + V'_{\mbox{eff}}(\phi) + 
\Pi\phi = 0 \label{eq:phi}
\end{equation}
where $' \equiv \partial/\partial \phi$. 
$\Pi$ in eq. (\ref{eq:phi}) generically represents
the backreaction of quantum fluctuations on the zero mode evolution via 
a change to the effective mass of the inflaton. It may be written 
specifically  as the  polarisation operator \cite{linde90}
or given explicitly in certain approximations,  such as  Hartree 
factorisation \cite{devegads}, or the large-$N$ 
limit of $O(N)$ vector models \cite{Boyan}.

The geometry of space (if the background is {\sc flrw})
enters only through the expansion $\Theta$, whose evolution 
is given by the Raychaudhuri equation \cite{elli71}, which in flat 
{\sc flrw} spacetime is: 
\begin{equation}
\dot{\Theta} = -\frac{3 \kappa}{2}\dot{\phi}^2
\label{eq:ray}  
\end{equation} 
Here $\kappa = 8\pi$ is the gravitational coupling constant.

To illustrate preheating and hence the first axis of the correspondence 
between reheating and gravitational 
wave evolution, assume that  $\phi$ interacts  with a light scalar field
$\chi$, which itself has no self-interaction, via the Lagrangian interaction 
term $\frac{1}{2}g^2\phi^2 \chi^2$.  Consider  the simplest effective 
potential for chaotic inflation: 
\begin{equation}
V(\phi) = \frac{1}{2}m_{\phi}^2 \phi^2\,.
\label{eq:quadpot}
\end{equation}
The solution of the inflaton equation of motion (\ref{eq:phi}), when the
frequency is an adiabatic invariant, is that of
decaying  sinusoidal oscillations, $\phi(t) = \Phi(t) \sin(m_{\phi}
t)$.  In the absence of particle production the amplitude varies roughly as 
$\Phi \sim \frac{1}{m_{\phi} t}$, due to the averaged expansion. The  time  
evolution of
the quantum fluctuations for each mode of the $\chi$-field  is
given by (neglecting $m_{\chi}$) \cite{KLS94}:
\begin{equation}  
\ddot{\chi}_k + \Theta \dot{\chi}_k + \left(\frac{k^2}{S(t)^2} + g^2 \Phi^2
\sin^2(m_{\phi} t) \right) \chi_k = 0
\label{eq:chi}
\end{equation}

This can be put in canonical Mathieu form if one neglects the expansion of 
the universe ($\Theta = 0$): 
\begin{equation}
\chi_k'' + \left[A(k) - 2q \cos(2 z)\right] \chi_k = 0
\label{eq:math}
\end{equation}
with dimensionless coefficients:
\begin{equation}
A_{\chi}(k) = \frac{k^2}{m_{\phi}^2 S^2} + 2q~~,~~~~ q_{\chi} = 
\frac{g^2  
\Phi^2}{4m_{\phi}^2}
\label{eq:mathparam}
\end{equation}
and  $z = m_{\phi} t$.
Depending on the size of the
coupling, $g$, and the mass $m_{\phi}$, {\em certain} modes $\chi_k$ will 
thus be amplified exponentially: $\chi_k = p_k(m_{\phi} t) 
\exp(\mu_{k}{}^{(n)} m_{\phi} t)$, where $p_k$ are functions with the 
same period as the oscillations of the inflaton field and the positive 
$\mu_k{}^{(n)}$ are the Floquet indices corresponding to the $n$-th 
instability band. The existence of resonance bands survives 
when the expansion of the universe is included \cite{Kaiser,devegads}.
The parameter $q$ divides the phase space into three broad classes with
qualitatively different behaviour. The case $q \ll 1$ is well understood
\cite{TYB94} and can be treated perturbitively, the effects of 
expansion  being important. The broad resonance case is described 
roughly by $\pi^{-1} < q < q_*~,~ q_* \sim 10^2$, is non-perturbative and 
requires  consideration of the backreaction of created particles on the 
zero mode  $\phi_0$. The resonance bands are characterized by huge occupation
numbers of produced particles, typically of order $n_k \sim \frac{1}{g^2}$. 
The upper limit, $q_*$, is for an expanding universe and is much lower
in Minkowski spacetime ($\Theta = 0$) \cite{PR96}. The wide resonance, $q 
\gg q_*$, evolution  is dominated by scattering effects which rapidly 
shut-off the exponential growth of the $\chi$ fluctuations 
\cite{tkachev,PR96}.

\section{Gravitational waves in Maxwell-Weyl form}\label{sec:tensor}

Gravitational waves (tensor perturbations) describe curvature fluctuations
unassociated directly with matter.  We will discuss their evolution using 
the elegant  covariant  formalism \cite{elli71}. 

The  covariant derivative of the
four velocity $u_a$ is split into kinematic parts via:
\begin{equation}
u_{a;b} = \sigma_{ab} + \omega_{ab} + \frac{1}{3} \Theta h_{ab} -
\dot{u}_a u_b
\label{eq:covfour}
\end{equation}
where the shear $\sigma_{ab} = \sigma_{(ab)}$ is traceless, 
$\omega_{ab} = \omega_{[ab]}$ is the vorticity and $h_{ab}
= g_{ab} + u_a u_b$ is both the projection tensor into, and the 
metric of, the hypersurfaces orthogonal to $u_a$ in the case where 
$\omega_{ab} = 0$ \footnote{This is always true for scalar fields with 
our choice of four velocity, since spatial gradients of a scalar commute.}. 
$\dot{u}_a$ is the acceleration and is caused by 
pressure gradients. Here indices surrounded by round brackets denote 
symmetrization on those indices while square brackets denote 
anti-symmetrization.

\subsection{The Electric and
Magnetic Weyl Tensors}

A purely tensor description of gravitational  waves is still partially 
lacking in the covariant approach \cite{roy}, but a sufficient
description was given first by Hawking (1966) \cite{hawk66} in terms 
of the magnetic $H_{ab}$  {\em  and} electric, $E_{ab}$ parts of the Weyl 
tensor, which are
automatically traceless and gauge-invariant (since they vanish
in exactly {\sc flrw} spacetimes) \cite{DBE96,elli71}, given by:
\begin{eqnarray}
H_{ac} &=& \frac{1}{2} \eta_{ab}{}^{gh} C_{ghcd} u^b u^d ~~ \equiv 
{}^*\!C_{abcd} u^b u^d \\ E_{ac} &=&  C_{abcd} u^b u^d 
\label{eq:defhe}
\end{eqnarray}
where $*$ is the usual dual (Hodge)
operator. These definitions are completely analogous to those of the  
electric and magnetic fields in terms of the field strength, $F_{a b}$, 
in standard electromagnetism \cite{EH96}. A
natural interpretation of the electric Weyl part is given by
the geodesic deviation equations, which for the special case of a plane 
gravitational wave propagating along the $x^1$ direction  are:
\cite{deC}
\begin{equation}
\ddot{\xi}^{\alpha} = E^{\alpha}{}_{\beta} \xi^{\beta}~~,~~~\alpha, \beta
= 2,3
\label{eq:geodev}
\end{equation}
where $\xi^{\alpha}$ is the connecting vector between orthonormal tetrads
associated with the congruence of null geodesics ruled by gravitons. 
This means that we can directly attribute the physical effects of linear
tensor perturbations, on e.g. a gravitational wave detector, to the electric
part of the Weyl tensor.

The evolution equations for $H_{ab}$ and $E_{ab}$ are provided by the
Bianchi identities:
\begin{equation} 
R_{ab[cd;e]} = 0  \Longleftrightarrow
C^{abcd}{}_{;d} = R^{c[a;b]} - \frac{1}{6} g^{c[a} R^{;b]}~,
\label{eq:bianchi}
\end{equation}
with the above equivalence only holding in four dimensions.  These
yield the {\em nonlinear} gravitational analogues of the Maxwell equations
\cite{elli71} which take the form of two evolution equations, $\dot{E}_{ab},
\dot{H}_{ab}$ and two constraint equations, $\mbox{div} E_{ab},~ \mbox{div}  
H_{ab} $ - see equations (\ref{eq:edot}), (\ref{eq:hdot}),
(\ref{eq:dive}), (\ref{eq:divh}), again just as in the case of Maxwell's 
equations.

In addition we can associate a natural super-energy to  
gravitational waves in the covariant approach through the scalar:
\begin{equation}
E_{ab} E^{ab} + H_{ab} H^{ab}~,
\label{eq:energy}
\end{equation}
which is in fact just the $u_a$-projected timelike component of the
Bel-Robinson tensor, the super-stress-tensor for the gravitational field 
\cite{zakh73,mash96}.  After expansion in eigenfunctions of
the tensor Helmholtz equation, and using the relation
$\nabla^2 = -k^2/S^2$,  the modes $H_k$, of the
magnetic part $H_{ab}$ satisfy (see \cite{DBE96} and appendix A  for the 
derivation): 
\begin{equation}
\ddot{H}_k + \frac{7}{3}\Theta \dot{H}_k + \left[
\frac{k^2}{S^2} +  \dot{\Theta} +
\Theta^2 + \frac{1}{2}(\mu - p)\right] H_k = 0~,
\label{eq:heq}
\end{equation}
where $\mu$ and $p$ are the relativistic energy density and pressure
respectively. This is a simple decoupled  equation (c.f. eq.
\ref{eq:chi}), while the shear satisfies:
\begin{equation}
\ddot{\sigma}_k + \frac{5}{3} \Theta \dot{\sigma}_k +
\left[\frac{k^2}{S^2} + \frac{\Theta^2}{9} +
\frac{1}{6}\mu - \frac{3}{2} p\right] \sigma_k = 0~.
\label{eq:shear}
\end{equation}
Finally the modes of the electric part obey an equation which is coupled
to the shear (c.f. equation \ref{eq:edot}):
\begin{eqnarray}
\ddot{E}_k &+& \frac{7}{3} \Theta \dot{E}_k + \left[\frac{k^2}{S^2} +
\dot{\Theta} +
\Theta^2 + \frac{1}{2}(\mu - p) \right] E_k \nonumber \\ &=& 
-\left[\frac{1}{3}\Theta(\mu + 
p) + \frac{1}{2}(\dot{\mu} + \dot{p})\right] \sigma_k
\label{eq:e}  
\end{eqnarray}
The shear can be eliminated from the
$E_k$ equation by another differentiation \cite{DBE96}, but since the shear 
is the variable which appears in the
covariant formula for the temperature anisotropies of the CMB, it will be
convenient to retain it explicitly.
The shear does not appear in the $E_{ab}$ equation  when
$\mu + p = 0$, i.e. in exact de Sitter
spacetime or in vacuum, be it Minkowski ($\Theta = 0$) or Milne ($\Theta 
\not = 0$). In this case there is an
exact symmetry between the equations (\ref{eq:heq},\ref{eq:e}) under the 
interchange 
$E_{ab} \leftrightarrow H_{ab}$. This is the linearised version 
of the exact nonlinear rotational duality that exists in the 
full vacuum Bianchi identities  \cite{MB97} and which is 
the gravitational
analogue of the electromagnetic duality in vacuum \cite{OM77} which
lead to the Montonen-Olive conjecture  and the modern progress in
dualities of supersymmetry and string theory  \cite{SW94}. See 
\cite{EH96,bonnor} for  further analysis of the  
gravity-electromagnetic duality in cosmology.

\subsubsection{Oscillatory dynamics in reheating}

Now let us specialise  equations (\ref{eq:heq}-\ref{eq:e})
to the case of classical scalar field dynamics.   
Treating here only the case of a single scalar field, 
we have, using the equivalence of $\phi$ with a perfect fluid:
\begin{equation}
\mu = \kappa \left[\frac{1}{2} \dot{\phi}^2 + V(\phi) \right]~,~~
p = \kappa \left[\frac{1}{2} \dot{\phi}^2 - V(\phi)\right]
\label{eq:enpres}
\end{equation}
so that
\begin{equation}
\frac{1}{2}(\mu - p) = \kappa V(\phi),~~ (\mu + p) = \kappa
\dot{\phi}^2~,~~  (\dot{\mu} +
\dot{p}) = 2 \kappa \dot{\phi} \ddot{\phi}~.
\end{equation}  
Here $V(\phi)$ is the effective potential of the scalar field.  Note that if 
scalar perturbations become important 
then these relations will gather terms proportional to $(\nabla \phi)^2$ 
\cite{KT}. This is precisely the case if one wishes to consistently study 
the production of tensor perturbations  from gravitational 
bremstrahlung \cite{tkachevgw}. However, it brings with it a host of 
complexities, since for example,  eq.s (\ref{eq:heq} - 
\ref{eq:e}) must be rederived in the presence of the backreaction of 
matter fluctuations, a highly non-trivial problem. We will not consider 
this issue further here.

The important point is that with the above
identifications, the equations (\ref{eq:heq}), (\ref{eq:shear}),
(\ref{eq:e}) become generalizations of
eq.(\ref{eq:chi}). There thus exists a strong connection
between the evolution governing reheating  under a given
potential and the equations governing the evolution $H_k, E_k$ and $\sigma_k$
particularly when $V(\phi)$ is an even polynomial in $\phi$, as in 
chaotic inflation.

\section{Chaotic inflation and  Duality} 

Consider again the quadratic chaotic  potential, Eq. (\ref{eq:quadpot}). In 
addition to chaotic inflation, this describes the  dynamics of the  invisible
axion, and the Polonyi and moduli fields of supergravity and string
theory, with appropriate changes to the values of the masses, $m^2 \equiv
V''(\phi)$.

Here we will neglect the expansion of the universe as before to delineate 
the duality \cite{Kaiser}.  This would of course not be adequate for the 
study of long-time oscillatory amplification, but is justified if the 
period of 
oscillations of $\phi$ is small compared with typical averaged expansion 
times, i.e. $m_{\phi} \gg \overline{\Theta}$.  However, the oscillations 
of the expansion rate should also be included in a full description 
\footnote{That  the expansion also oscillates can be seen from eq. 
(\ref{eq:ray}).}\cite{pretend2}  which  will act
as an additional source of resonance. This will be important in obtaining 
the tensor spectrum from oscillating cold dark matter (CDM) relics like the 
invisible axion \cite{kim}. The equation (\ref{eq:heq}) 
for the modes, $H_k$, of the magnetic part $H_{ab}$ now becomes:
\begin{equation}
\ddot{H}_k + \left(\frac{k^2}{S^2} +
\frac{\kappa m_{\phi}^2}{2}\Phi^2 \sin^2(m_{\phi}t) \right) H_k = 0
\label{eq:hquad} \end{equation}

\begin{figure}
\epsfxsize = 2in
\epsffile{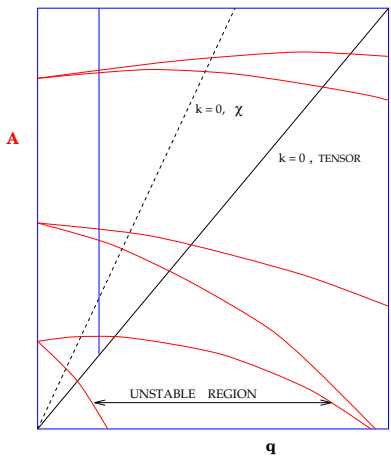}
\vspace*{0.5cm}
\caption{A schematic of the instability chart for the Mathieu equation. The 
two diagonal lines correspond respectively to the zero-modes ($k = 0$) of 
the  gravitational wave spectrum and the $\chi$-spectrum of eq. (4),
for the case \protect{$g^2 <  m_{\phi}^2$}. The vertical line near the 
left border of the chart is a 
sample spectrum for a given model (i.e. value of \protect{ $q$}).}
\label{fig:1}
\end{figure}  

This is the Mathieu equation and is  precisely the same as equation
(\ref{eq:chi}) for the evolution of the quantum fluctuations $\chi_k$,
with the replacement $g^2 \rightarrow
\kappa \frac{m^2_{\phi}}{2} = 4\pi m^2_{\phi}$. Hence $q_H = \pi \Phi$.
The requirement that production of $\chi$ bosons is more efficient than   
$H_{ab}$ amplification is then $g^2 > 4 \pi m_{\phi}^2$. If $m_{\phi}
\sim 10^{-6}$, as required to match CMB observations \cite{salopek}, then 
this is a weak constraint on the coupling $g$, namely $g^2 > 10^{-11}$.
Nevertheless, it is a constraint independent of $\Phi$ and hence applies 
to both chaotic and new inflationary models (which have quadratic 
potentials near the global minima). If the constraint is not met, it implies 
that reheating occurs preferentially via production of gravitons rather 
than the $\chi$-channel \footnote{This is not the only constraint to be 
met however. If the $\chi$ field has moderate or strong 
self-interactions,  the $\chi$-resonance is strongly suppressed 
\cite{PR96}. This is also the  case if $q_{\chi} \gg 10^3$, due to 
rescattering effects \cite{PR96,tkachev}. In  these cases, the graviton 
decay-channel could still be important and perhaps even dominant.}.

In figure 1 this situation is depicted schematically on the instability   
chart of the Mathieu equation: namely, the situation in which $q_H > 
q_{\chi}$. The two diagonal lines corresponding to the mode $k = 0$, 
delimit the physical (i.e. upper) region of  the chart. The single 
vertical line corresponds to a sample spectra for a
given value of $q$. Note that for this value of $q$, the tensor spectrum has
modes in the  first fundamental resonance band, while there are no $\chi$
modes which lie in this band. Figure (2) shows a  numerical 
integration of the spectrum as a function of time for modes in both 
stable and unstable bands. Figure (3) zooms in on the spectrum within a  
stable band.

Can $H_{ab}$ be amplified in the broad resonance regime ? This is 
necessary if gravitational wave enhancement is to be really effective 
over the expansion. This  requires   
$q_H > \pi^{-1}$. Since $q_H = \pi \Phi^2$ this implies $\Phi > \pi^{-1}$
in units of the Planck energy. In the case of chaotic
inflation, the amplitude of oscillations  goes as \cite{linderice} 
$\Phi \sim 1/N$, where $N$ is the number of oscillations of $\phi$, 
neglecting the non-equilibrium backreaction at the end of preheating 
which often leads to a sudden decrease in $\Phi$ \cite{Boyan}.
Thus during preheating proper, $q_{H} \sim \pi/N^2$, and $H_{ab}$ is 
initially  amplified in  the broad  resonance regime, but moves 
rapidly towards narrow resonance.

\begin{center}  
\begin{tabular}{|c|c|c|c|}
 \hline
 \hline
~~~    &  $\chi_k$  & $H_k$ &   $\sigma_k$ \\
\hline
\hline
& & & \\
 ~~~  A~~~ &~~~ $\frac{k^2}{S^2 m_{\phi}^2} + 2q$~~~  &~~~ 
$\frac{k^2}{S^2 m_{\phi}^2} + 2q$~~~  &~~~ $\frac{k^2}{S^2 m_{\phi}^2} + 
\frac{2}{19} q$~~~  \\

& & & \\

~~~ q ~~~&~~~ $\frac{g^2 \Phi^2}{4 m_{\phi}^2}$~~~ &~~~ $\pi \Phi^2$  &
$\frac{19}{6} \pi \Phi^2$~~~ \\
& & & \\
\hline \hline
\end{tabular}
\end{center}

Consider now the equations for the electric Weyl field, $E_{ab}$, and 
the shear $\sigma_{ab}$. They form a
partially decoupled linear system with time-dependent coefficients that 
form an approximately Floquet system. Here we ignore the 
decay of the amplitude. This is appropriate during preheating while  the 
non-equilibrium 
effects  and backreaction of graviton production are not too strong and 
the oscillation period of $\phi$, (controlled by 
$m_{\phi}$) is short compared with the expansion time scale. This gives:
\begin{equation}
\ddot{E}_k + \left(\frac{k^2}{S^2} + \frac{m_{\phi}}{2}\Phi^2
\sin(m_{\phi} t) \right)
E_k =  \frac{1}{8}\left[ \Phi^2 m_{\phi}^3 \sin(2m_{\phi} t) \right]  
\sigma_k 
\label{eq:equad }
\end{equation}
with
\begin{equation}
\ddot{\sigma}_k + \left( \frac{k^2}{S^2} - \kappa \frac{3}{4}
\dot{\phi}^2 + \kappa \frac{5}{6}m_{\phi}^2 \phi^2  \right)\sigma_k = 0
\label{eq:shearquad}
\end{equation}
The equation for the shear can again be cast in the form of the
Mathieu equation with 
\begin{equation}
A(k)_{\sigma} = \frac{k^2}{m_{\phi}^2 S^2} +
\frac{2}{19}q_{\sigma}~~, ~~~q_{\sigma} = \kappa \frac{19}{48} \Phi^2~.
\end{equation} 
A  comparison of parameters is given in table 1.

Note that the shear {\em always} lies in a region of broader resonance 
than the magnetic part of the Weyl tensor because $q_{\sigma} > 
q_{H}$ and $A_{\sigma}(k) < A_H(k)$. Now $q_{\sigma} = \frac{19 \pi}{6}
\Phi^2$ with the requirement of broad resonance amplification of the shear
$\Phi > \sqrt{\frac{6}{19}} \pi^{-1} \sim 0.56 \pi^{-1} \sim 0.18$.

Since it is the shear which directly
determines the
tensor contribution to the anisotropy  of the CMB, 
this may allow one to place constraints on large-amplitude  oscillatory
reheating. Since power-law inflation is known to produce one of the
strongest tensor signals during slow-roll \cite{lidsey} the signal to 
noise (due to cosmic variance and instrument) ratio for the tensor 
component in the CMB may be significantly larger than previously hoped 
\cite{knox}. The CMB anisotropy from a tensor signal is \cite{russ,dunsby}: 
\begin{equation}
\left( \frac{\delta T}{T} \right)^* = -\int_E^R   S~ \sigma_{ab}~ k^a
k^b d\lambda
\label{eq:covcmb}
\end{equation}
The left hand side is a gauge-invariant measure of the anisotropy in the
CMB, and $k^a$ is the wave vector ruling our past null cone. 

Now for purely tensor
perturbations,  $\sigma_{ab}$ and $E_{ab}$ are related by \cite{DBE96}:
\begin{equation}
\dot{\sigma}_{ab} = - \frac{2}{3} \Theta \sigma_{ab} - E_{ab}
\label{eq:shev}
\end{equation}
we see that exponential growth of the shear implies exponential growth of
the electric Weyl field, and hence the energy in gravitational
waves, $\Omega_{GW}$, increases  exponentially, via eq. (\ref{eq:energy}).

\begin{figure}
\epsfxsize=3in
\epsffile{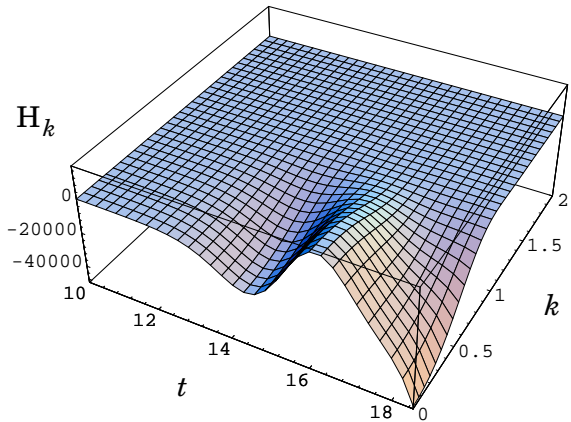}
 \caption{The gravitational wave spectrum, $H_k$ as a   
 function of
 $k$ and $t$, with the initial condition $H(0) = 0, \dot{H}(0) = 
10^{-4}$ and starting in the broad resonance regime. As can be 
seen, the power spectrum becomes highly $k$-dependent, breaking 
the usual scale-invariance of
the envelope predicted by inflation. Further, the  amplitude of the 
spectrum is
exponentially enhanced by resonant reheating over its value during 
inflation.}
\label{fig:2} 
\end{figure}
\begin{figure}
\epsfxsize=3in
\epsffile{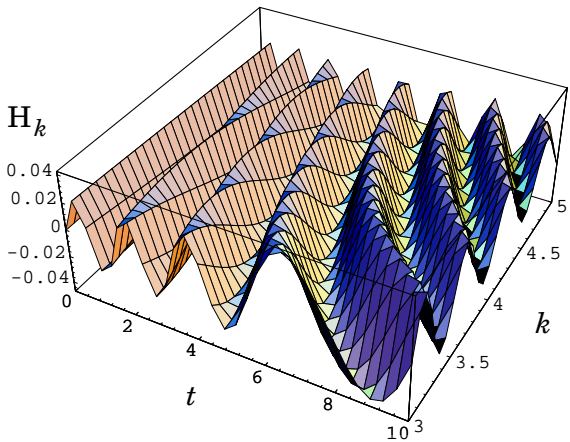}
\caption{A zoom of the stable band region  $3 \leq k \leq 5$. 
The solution is the usual bounded oscillatory one.  Note the scale on the 
$z$-axis (c.f. fig. 2).} \label{fig:3} \end{figure}


However, a counter example to the above situation occurs 
if there is non-thermal symmetry 
restoration \cite{KLS95,Boyan}: start with $\phi \gg M_{pl}$, 
as in the chaotic inflation scenario, but with a Coleman-Weinberg type 
new-inflationary potential, which is flat near the origin. If preheating 
restores symmetry, a phase of new inflation begins  with $\phi = 
0$. This redshifts away the resonantly amplified tensor spectrum. The 
second stage of reheating will be much less effective in amplifying the 
tensor spectrum than the first if $V(0) \ll M_{pl}^4$. This is the 
generic case for non-thermal symmetry restoration and hence   
amplification will occur in the very narrow resonance region where 
expansion  effects are expected to dominate, leaving an almost untouched, 
scale-invariant tensor spectrum.

\section{Conclusions}

The main result of this paper is that gravitational wave perturbations 
can be naturally amplified during a second order phase transition. Thus, in 
addition to  the standard creation of a scale-invariant stochastic 
gravitational wave 
spectrum  due to quantum fluctuations during inflation, there 
may be amplification of this spectrum during reheating which {\it breaks} 
the scale invariance and may significantly enhance the {\em rms} 
amplitude of the tensor perturbation spectrum with no {\em a priori} 
limit on the maximum wavelength affected. However the exact 
pattern and size of this ``symmetry breaking" is highly model dependent, 
and is hence relevant to inflationary potential reconstruction 
attempts \cite{lidsey,knox}. This amplification  is qualitatively different 
from gravitational  bremstrahlung \cite{tkachevgw} since it is due to 
the coherent oscillation  of the {\em mean} energy density and pressure of 
the inflaton, and is largely insensitive to the nature of the scalar 
field fluctuations. 

We have further examined  chaotic inflation with a quadratic potential in 
detail and  found  a duality between the equations describing the growth 
of $\chi$ 
fluctuations and those of the magnetic part of the Weyl tensor and the 
shear, the latter being  the variable which determines the 
gravitational wave 
signal in the CMB; eq (\ref{eq:covcmb}). Finally, the resonant 
enhancement of 
the tensor spectrum is completely missing  in those models of 
inflation which involve a continuous production of entropy during inflation, 
such as in 
the ``warm" inflation models \cite{warjun}. Future work will examine the 
quantitative importance of expansion, quantum backreaction and scattering 
effects  for the amplification of gravitational waves, and will compare 
the Weyl formalism with the more typical Bardeen approach.

\section*{Acknowldegements}
  
It is a pleasure to thank George Ellis, Andre Linde, Claudio Scrucca, 
Stefano Liberati, Roy Maartens and Peter Dunsby for  discussions and 
Andre Linde for  pointing out that the idea of resonant gravity wave 
amplification due to the oscillation of the expansion had  been 
worked on, though  previously unpublished. 

\appendix 
\section{Derivation of the wave equations}

The Bianchi identities, eq. (\ref{eq:bianchi}), yield the following
evolution equations \cite{DBE96}:

\begin{eqnarray}
h^m{}_ah^t{}_c \dot{E}^{ac} +
J^{mt} - 2H_a{}^{(t} \eta^{m)bpq}u_b \dot{u}_p + h^{mt}\sigma^{ab}E_{ab}
+\nonumber \\
+ \Theta E^{mt} - 3E_s{}^{(m}\sigma^{t)s} - E_s{}^{(m}\omega^{t)s} = -
\frac{1}{2}(\mu + p)\sigma^{tm}\,,
\label{eq:edot}
\end{eqnarray}
the `$\dot{E}$' equation, $h_{ab} = g_{ab} + u_a u_b$ is  the
projection tensor into the hypersurfaces orthogonal to  
$u_a$, and
\begin{eqnarray}
h^m{}_ah^t{}_c \dot{H}^{ac} -
I^{mt} + 2E_a{}^{(t} \eta^{m)bpq}u_b \dot{u}_p + h^{mt}\sigma^{ab}H_{ab}
+\nonumber \\
+ \Theta H^{mt} - 3H_s{}^{(m}\sigma^{t)s} - H_s{}^{(m}\omega^{t)s} = 0
\label{eq:hdot}
\end{eqnarray}
is the `$\dot{H}$' equation, where  we have given only the perfect fluid
form, and where the covariant curl terms are defined as:
\begin{eqnarray}
J^{mt} &=& h_a{}^{(m} \eta^{t)rsd}u_r H^a{}_{s;d}\\
I^{mt} &=& h_a{}^{(m} \eta^{t)rsd}u_r E^a{}_{s;d}
\label{eq:curl}
\end{eqnarray}
These are the fully nonlinear equations and despite their complexity, note
the symmetries between the equations (\ref{eq:edot}) and (\ref{eq:hdot}),
broken only by the driving term $-\frac{1}{2}(\mu + p)\sigma_{ab}$ in eq.
(\ref{eq:edot}) which is absent in the $\dot{H}_{ab}$ equation.
In addition there are the gravitational `div E' and `div H' constraint
equations: 
\begin{equation}
h^t{}_a E^{as}{}_{;d}h^d{}_s - \eta^{tbpq}u_b\sigma^d{}_pH_{qd} + 3
H^t{}_s \omega^s = \frac{1}{3} h^{tb} \mu_{;b} \,, 
\label{eq:dive}
\end{equation}
\begin{equation}
h^t{}_a H^{as}{}_{;d}h^d{}_s + \eta^{tbpq}u_b\sigma^d{}_pE_{qd} - 3
E^t{}_s \omega^s = (\mu + p)\omega^t\,,
\label{eq:divh}
\end{equation}
where the vorticity vector is $\omega^a  = \frac{1}{2} \eta^{abcd} u_b
\omega_{cd}$.

In linear theory about a {\sc flrw} background, to specify that we are 
interested in purely tensor, i.e. gravitational
wave solutions, we require that there are no scalar or vector perturbations:
\begin{equation}
\frac{1}{3} h_a{}^b \mu_{;b} = \omega_a = 0
\label{eq:lindiv}
\end{equation}
Examining equations (\ref{eq:dive},\ref{eq:divh}) and dropping all second
order terms, this implies that the divergences themselves vanish:
\begin{equation}
h^t{}_a E^{as}{}_{;d}h^d{}_s = h^t{}_a H^{as}{}_{;d}h^d{}_s = 0
\end{equation}
This ensures only purely  radiative solutions,
and since $E^a{}_a = H^a{}_a = 0$, by the construction
of the Weyl tensor, we have the analogue of the transverse-traceless (TT)
conditions usually imposed on tensor metric perturbations.

The equations
(\ref{eq:edot},\ref{eq:hdot}) can be linearised consistently \cite{DBE96},
giving the coupled set:
\begin{equation}
\dot{E}_{ab} + \Theta E_{ab} + h_{(a}^f \eta_{b)cde} u^c \nabla^e H_f^d +
\frac{1}{2} (\mu + p) \sigma_{ab} = 0
\label{eq:almoste}
\end{equation}
and
\begin{equation}
\dot{H}_{ab} + \Theta H_{ab} - h_{(a}^f \eta_{b)cde} u^c \nabla^e E_f^d =
0
\label{eq:almosth}
\end{equation}  
These can be converted to wave equations for $E_{ab}$ and $H_{ab}$ by
differentiation (see \cite{DBE96}). Similarly the shear,  $\sigma_{ab}$,
satisfies a wave equation and the three variables form  a partially - 
coupled system. Once one transforms the variables to momentum space via 
expansion in eigenfunctions, $Q^k_{ab}$, of the tensor  Helmholtz 
equation, one obtains the ordinary differential equations 
(\ref{eq:heq}), (\ref{eq:shear}), (\ref{eq:e}).


\begin{references}

\bibitem{DL} A.D. Dolgov and A.D. Linde, Phys. Lett. {\bf 116B},  329
(1982);   L.F. Abbott, E. Fahri, and M. Wise,  Phys. Lett. {\bf
117B}, 29 (1982).

\bibitem{Brand}  J. Traschen and R. Brandenberger,   Phys. Rev. D {\bf
42},  2491 (1990); A.D. Dolgov and D.P. Kirilova,  Sov. Nucl. Phys.
{\bf 51},  273 (1990).

\bibitem{KLS94} L. Kofman, A. Linde, and A. A.
Starobinsky,  {\it Phys. Rev. Lett.}  {\bf 73}, 3195 (1994)

\bibitem{TYB94} Y. Shtanov, J. Traschen  and R. Brandenberger,  Phys. Rev. 
D  {\bf 51}, 5438 (1995) 

\bibitem{KKT94} M. Kamionkowski, A. Kosowsky, and M. S. Turner, Phys. 
Rev. D {\bf 49}, 2837 (1994); astro-ph/9310044 

\bibitem{tkachevgw}  S. Yu. Khlebnikov, I. I. Tkachev,  
hep-ph/9701423 (1997)



\bibitem{hawk66} S. W. Hawking,  Ap. J. {\bf 145}, 544 (1966)

\bibitem{elli71} G. F. R. Ellis, {\em Relativistic Cosmology}, 
in {\em Carg\'ese  Lectures in
Physics}, vol. VI, ed. E. Schatzmann (Gordon and Breach, 1973), p.1

\bibitem{BED92} M. Bruni, P. K. S. Dunsby and G. F. R. Ellis, Class. 
Quant. Grav. {\bf 9}, 921  (1992)

\bibitem{gris} L. P. Grishchuk, Phys. Rev. D {\bf 53}, 6784, 
(1996)~;~{\em ibid} {\bf 50}, 7154 (1994)

\bibitem{deru} N. Deruelle and V. F. Mukhanov, Phys. Rev. D {\bf 52}, 5549
(1995)

\bibitem{martin} J. Martin and D. J. Schwarz,  gr-qc/9704049 
(1997) 

\bibitem{DBE96}  P. K. S. Dunsby, B. A. Bassett and G. F. R. Ellis, Class.
Quant. Grav., {\bf 14}, 1215  (1997)  

\bibitem{KLS95} L. Kofman, A. Linde and A. A. Starobinsky, {\it Phys. 
Rev. Lett.} {\bf 76}, 1011  (1996)

\bibitem{devegads} D. Boyanovsky, D. Cormier, H.J. de Vega, R. Holman, 
Phys. Rev. D {\bf 55}, 3373 (1997),  D. Boyanovsky, D. Cormier, H. J. de 
Vega, R. Holman, A. Singh, M. Srednicki, hep-ph/9703327 (1997)


\bibitem{Boyan}  D. Boyanovsky, H. J. de Vega, R. Holman, J. F. J. 
Salgado,  Phys.Rev. D {\bf 54}, 7570 (1996);
~ D. Boyanovsky, H.J. de Vega, R. Holman, D.S. Lee, and
A. Singh, {\it Phys.Rev. D} {\bf 51}, 4419 (1995);~  D. Boyanovsky, M. 
D'Attanasio, H. J. de Vega, R. Holman, D.-S. Lee, and A. Singh,    
Phys. Rev. D {\bf 52}, 6805 (1995)

\bibitem{Yoshimura} M. Yoshimura, Prog. Theo. Phys. {\bf 94}, 873 
(1995); H. Fujisaki, K. Kumekawa, M. 
Yamaguchi, and M. Yoshimura, Phys. Rev. D {\bf 53}, 6805 (1995)

\bibitem{pretend2} B. A. Bassett, {\em in preparation}  ~(1997)

\bibitem{Kaiser} D. Kaiser, Phys. Rev. D {\bf 53}, 1776 
(1996); D. Kaiser, hep-ph/9702244 (1997)

\bibitem{tkachev} S. Yu. Khlebnikov, I. I. Tkachev,
 hep-ph/9610477 and hep-ph/9608458, (1996) and S. Yu. Khlebnikov, I. I. 
Tkachev, Phys. Rev. Lett. {\bf 77}, 219 (1996)

\bibitem{PR96}  T. Prokopec, T. G. Roos, hep-ph/9610400 (1996) 

\bibitem{EH96}  G. F. R. Ellis and P. Hogan, Gen. Rel. Grav. {\bf
29}, 235 (1996)

\bibitem{bonnor} W. B. Bonnor,  Class. Quant. Grav. {\bf  12},  No 2, 499
(1995), \& {\em ibid}, {\bf 12} No 6, 1483 (1995)

\bibitem{deC}  F. de Felice and C. J. S. Clarke {\em Relativity on
Curved Manifolds}, (Cambridge Univ. Press,  1990)

\bibitem{zakh73} V. D. Zakharov, {\em Gravitational Waves in Einstein's
Theory}, Trans. R. N. Sen, (Halsted, New York, 1973).

\bibitem{mash96} B. Mashoon, J.C. McClune and H. Quevedo, Preprint, 
gr-qc/9609018 (1996)

\bibitem{KT}  E. W. Kolb and M. S.Turner, {\em The Early Universe} 
(Addison-Wesley, 1990)

\bibitem{kodama1}T. Hamazaki and H. Kodama, gr-qc/9609036 (1996); 
H. Kodama and T. Hamazaki, gr-qc/9608022 (1996)) 

\bibitem{NT96} Y. Nambu and A. Taruya, gr-qc/9609029 (1996)

\bibitem{ml96} A. Linde and V. Mukhanov, astro-ph/9610219 (1996)

\bibitem{kim} J.E. Kim, Physics Reports, {\bf 150}, 1 (1987)

\bibitem{salopek}  D.S. Salopek, Phys. Rev. D, {\bf 43}, 3214 (1991)

\bibitem{linderice} A. D. Linde,  in lectures of {\em the School 
for Astrofundamental Physics},  (Erice, 1996)

\bibitem{russ} H. Russ, M. Soffel, C. Xu and P.K.S. Dunsby, 
Phys. Rev. D, {\bf 48}, 4552 (1993)

\bibitem{dunsby} P. K. S. Dunsby, UCT Preprint, {\em A fully covariant 
description of CMB anisotropies },~ (1997)

\bibitem{roy}  R. Maartens, G. F. R. Ellis and S. T. C. Siklos,  
Class. Quant. Grav. {\bf 14},  (1997) gr-qc/9611003; R. 
Maartens, Phys. Rev. D {\bf 55}, 463 (1997)

\bibitem{linde90} A.D. Linde, {\it Particle Physics and Inflationary
Cosmology} (Harwood, Chur, Switzerland, 1990).

\bibitem{MB97}  R. Maartens and B. A. Bassett,  gr-qc/9704059 (1997)

\bibitem{OM77}  C. Montonen and D. I. Olive, Phys. Lett. {\bf 72 B}, 
117, (1977)

\bibitem{SW94}  N. Seiberg and E. Witten, Nucl. Phys. {\bf B 426}, 19 (1994);
{\em Erratum} {\bf B 430}, 485 (1994)

\bibitem{LS} D. H. Lyth and  E. D. Stewart, Phys. Rev. Lett. {\bf 75},
201 (1995); D. H. Lyth and E. D. Stewart,  Phys. Rev. D {\bf 53} 1784 (1996)

\bibitem{lidsey} J.E. Lidsey {\em et al}, astro-ph/9508078, (1995)

\bibitem{knox} L. Knox and M.S. Turner, Phys. Rev. Lett. {\bf 73}, 3347 
(1994)

\bibitem{warjun}  A. Berera, Phys. Rev. D {\bf 54}, 2519, (1996)

\end{references}
\end{document}